\documentclass[10pt,conference]{IEEEtran}
\IEEEoverridecommandlockouts

\usepackage[T1]{fontenc}
\usepackage{url}
\usepackage{ifthen}
\usepackage{cite}
\usepackage[cmex10]{amsmath} 
\usepackage{longtable}
\usepackage{makecell}
\usepackage{algorithm,algorithmicx}
\usepackage[noend]{algpseudocode}
\usepackage{color}
\usepackage[normalem]{ulem}
                             
\usepackage{amssymb,amsthm}
\usepackage{mathtools,bbm}  
\usepackage{xifthen,xparse}
\usepackage{hyperref}  
\usepackage{float}
\usepackage{hhline}
\usepackage{pgfplots}
\usepackage{caption}
\usepackage{subcaption}
\usepackage{rotating}

\usepackage{stmaryrd}

\usepackage{booktabs,tabularx}
\newcolumntype{C}{>{\centering\arraybackslash}X}

\usepackage[table]{xcolor}

\definecolor{dkgreen}{rgb}{0,0.6,0}
\definecolor{gray}{rgb}{0.5,0.5,0.5}
\definecolor{mauve}{rgb}{0.58,0,0.82}

\usepackage{pgfplots}
  \pgfplotsset{compat=newest}
  \usetikzlibrary{plotmarks}
  \usetikzlibrary{arrows.meta}
  \usepgfplotslibrary{patchplots}
  \usepackage{grffile}
  \usepackage{amsmath}

\usepackage{tikz,pgfplots,tkz-graph,color}
\pgfplotsset{compat=newest}
\pgfplotsset{plot coordinates/math parser=false}
\usetikzlibrary{decorations.markings,calc,positioning,matrix}
\usepackage{quantikz}
\usepackage{graphicx}

\usepackage{textcomp}

\allowdisplaybreaks

\newtheorem{theorem}{Theorem}
\newtheorem{corollary}[theorem]{Corollary}
\newtheorem{remark}[theorem]{Remark}
\newtheorem{lemma}[theorem]{Lemma}

\newtheorem{definition}{Definition}
\newtheorem{example}{Example}

\newenvironment{example*}
  {\addtocounter{example}{-1}\example}
  {\endexample}

\newif\ifnotes
\notestrue

\newcommand{\llbr}{[\![}
\newcommand{\rrbr}{]\!]}

\begin{document}

\title{Fault Tolerant Quantum Simulation\\ via Symplectic Transvections}

 

\author{\IEEEauthorblockN{Zhuangzhuang Chen}
\IEEEauthorblockA{\textit{Dept. of Elec. Comp. Engg.} \\
\textit{University of Arizona}\\
Tucson, USA \\
zhuangzhuangchen@arizona.edu}
\and
\IEEEauthorblockN{Jack Owen Weinberg}
\IEEEauthorblockA{\textit{Dept. of Elec. Comp. Engg.} \\
\textit{University of Arizona}\\
Tucson, USA \\
jackweinberg@arizona.edu}
\and
\IEEEauthorblockN{Narayanan Rengaswamy}
\hfill\IEEEauthorblockA{\textit{Dept. of Elec. Comp. Engg.} \\
\textit{University of Arizona}\\
Tucson, USA \\
narayananr@arizona.edu}
\thanks{The work of N.~R. was partially supported by the U.S. NSF under Grant no. 2106189. We thank Asit Kumar Pradhan for his help with simulations.}
}

{\maketitle}

\begin{abstract}
Conventional approaches to fault-tolerant quantum computing realize logical circuits gate-by-gate, synthesizing each gate independently on one or more code blocks. 
This incurs excess overhead and doesn't leverage common structures in quantum algorithms. 
In contrast, we propose a framework that enables the execution of \emph{entire logical circuit blocks at once}, preserving their global structure. 
This whole-block approach allows for the direct implementation of logical Trotter circuits—of arbitrary rotation angles—on any stabilizer code, providing a powerful new method for fault tolerant Hamiltonian simulation within a single code block. 
At the heart of our approach lies a deep structural correspondence between symplectic transvections and Trotter circuits. 
This connection enables both logical and physical circuits to share the Trotter structure while preserving stabilizer centralization and circuit symmetry even in the presence of non-Clifford rotations. 
We discuss potential approaches to fault tolerance via biased noise and code concatenation. 
While we illustrate the key principles using a $[[8,3,3]]$ code, our simulations show that the framework applies to Hamiltonian simulation on even good quantum LDPC codes. 
These results open the door to new algorithm-tailored, block-level strategies for fault tolerant  circuit design, especially in quantum simulation.
\end{abstract}

\begin{IEEEkeywords}
Trotter circuits, stabilizer codes, symplectic transvections, fault tolerance, Pauli propagation
\end{IEEEkeywords}

\section{Introduction}
\label{sec:introduction}

Quantum algorithms are expected to provide significant theoretical and practical advantages for specific, hard, computational problems such as factoring large integers or simulating the structure of complex molecules.
Since each hardware component, such as state preparation, gates, and measurement, are noisy, the algorithms cannot be implemented reliably on raw quantum data.
The most common approach towards fault tolerant execution of these algorithms is to encode the data in a quantum error correcting code (QECC) and synthesize noise-resilient realizations of the constituents of the algorithm's circuit.
The redundancy in the code enables one to detect and correct errors in such realizations, thereby preventing the noise from overwhelming the system.
These fault tolerant realizations must be designed carefully to avoid errors from spreading catastrophically through the circuits.

The conventional approach to QECC-based universal fault tolerant quantum computing is to construct codes with good parameters and demonstrate fault tolerant implementations of a universal set of logical gates, e.g., CNOT, Hadamard, and $T$.
Given such implementations, one can execute any quantum algorithm by combining the fault tolerant realizations of each component gate in the appropriate order.
The standard methods for these realizations are transversal or fold-transversal gates~\cite{Breuckmann-quantum24}, code automorphisms~\cite{Sayginel-arxiv24,Malcolm-arxiv25}, lattice surgery~\cite{Horsman-njp12}, code switching~\cite{Anderson-prl14}, gauge fixing~\cite{Vuillot-njp19}, and magic state distillation~\cite{Bravyi-pra05}.
However, such gate-by-gate approaches incur large overhead since they typically need multiple code blocks and rounds of error correction per logical gate to guarantee fault tolerance.
This fundamentally affects the run-time of the entire algorithm, thereby potentially defeating any quantum advantage. 

We disrupt this conventional wisdom by proposing to synthesize \emph{entire logical circuit blocks at once}.
Indeed, many quantum algorithms share common subroutines such as Trotter circuits in quantum simulation~\cite{Whitfield-molphy11,Daley-nature22}.
These are structured blocks in logical circuits that act on multiple qubits, which can be leveraged for fault tolerant synthesis.
Despite the quantum computer being universal, the compiler can make smart choices in executing different algorithms rather than taking a one-size-fits-all approach.
Such an algorithm-aware co-design of fault tolerance with quantum algorithms is severely underexplored in the literature.
In prior work~\cite{Chen-arxiv24,Chen-qce24}, we demonstrated optimal-depth flag-based fault tolerant execution of Clifford Trotter circuits on the error-detecting family of \emph{Iceberg codes}~\cite{Self-natphys24,Yamamoto-prr24}.
We introduced a \emph{solve-and-stitch} approach that synthesizes a logical Clifford Trotter circuit by solving for small circuits satisfying individual Pauli constraints and stitching them appropriately.
While this demonstrates the utility of synthesizing logical blocks as a whole, it is limited to Clifford Trotter circuits on specific error-detecting codes.

In this paper, we overcome these limitations and introduce an exciting new strategy to implement \emph{any logical Trotter circuit on a single code block of any stabilizer code}.
The synthesized circuit on the code qubits also shares the Trotter structure, commutes with all stabilizers, and has no constraints on the rotation angle in the logical circuit.
We achieve this by a new fundamental insight that relates Trotter circuits to symplectic transvections~\cite{Koenig-jmp14,Rengaswamy-tqe20} in the Clifford case and extends the connection to non-Clifford settings. 
We further show that the propagation of Pauli operators admits a clean linear decomposition and satisfies a novel double-angle exponential relation.
The approach can be directly applied to implement Hamiltonian simulation, or related algorithms, on a single code block of quantum low-density parity-check (QLDPC) codes.
We also discuss strategies to make the circuits fault tolerant, with potentially much less overhead than conventional methods such as combining lattice surgery with magic state distillation and injection.
There are several advantages: 
\begin{enumerate}

\item Targeted logical gates on specific logical qubits of a QLDPC code is challenging (there is progress~\cite{Patra-arxiv24}). The efficiency with which these codes pack logical qubits makes it difficult to address different logical qubits in an isolated manner. Our approach circumvents this issue by executing Trotter circuits that act on multiple logical qubits at once. The physical circuits in our synthesis will act on the support of some logical Pauli operator that depends on the logical Trotter circuit.

\item Transversal gates or lattice surgery involves multiple code blocks in not only topological codes, but also good QLDPC codes. This defeats the advantage provided by such codes that encode many logical qubits in a single code block. Our approach enables one to run sequences of Trotter circuits on one code block, up to some code concatenation that may be needed for fault tolerance.

\item In the conventional gate-by-gate approach, one must perform multiple rounds of error correction per logical gate to ensure fault tolerance. By focusing on non-trivial blocks of logical circuits, we reduce the frequency of error correction while enabling the tracking of errors through the realizations of these blocks.

\item Compilation techniques are typically developed only with physical circuits in mind, thereby not leveraging the connections between logical and physical circuits on a code. Since our realization of logical Trotter circuits preserves the Trotter structure even at the physical level, any compilation strategy developed in the NISQ era on non-error-corrected circuits can be directly applied even for fault tolerant execution of quantum simulation.

\item Decoders are typically developed without specific circuit structures in mind, except in certain cases such as magic state injection or multi-qubit Pauli measurements. This makes their design difficult as errors can propagate very differently in different circuit structures. The Trotter structure in our physical circuits enables exciting new possibilities for designing syndrome extraction and decoding that is tailored to the algorithm. We think this will lead to resource reduction and improved thresholds.

\item Recent innovations on algorithmic fault tolerance~\cite{Zhou-arxiv24} can be integrated into our approach. For example, since quantum algorithms end with measurements, which can be assumed in general to be $Z$-measurements, it suffices to suppress logical $X$ errors at the end of all unitary gates. This insight can be combined with biased noise and bias-preserving gates to enable a fundamentally new framework for fault tolerant quantum simulation.

\end{enumerate}

The paper is organized as follows.
Section~\ref{sec:main_results} discusses our key insights and results,
Section~\ref{sec:background} introduces the necessary technical background, Section~\ref{sec:clifford_trotter} reveals the deep connection between symplectic transvections and Clifford Trotter circuits, Section~\ref{sec:non_clifford_trotter} demonstrates the extension to arbitrary Trotter circuits, Section~\ref{sec:fault_tolerance} explores few different approaches to fault tolerance, and Section~\ref{sec:conclusion} concludes the paper.  

\section{Main Results}
\label{sec:main_results}

Trotter circuits are exponentiated $m$-qubit Pauli operators, $\exp\left( -\imath \frac{\theta}{2} P \right)$, where $\imath = \sqrt{-1}$ and $P$ is Hermitian.
These form the building blocks of many quantum algorithms, specifically for quantum simulation via the Trotter-Suzuki decomposition (which we discuss later).
Such algorithms executed a sequence of Trotter circuits corresponding to different Pauli operators, $P$, and rotation angles, $\theta$, depending on the specific Hamiltonian and problem description.
Hence, if we optimize the fault tolerant execution of these circuits, then we can execute them as many times as necessary.

Our fundamental insight is that for $\theta = \frac{\pi}{2}$, Trotter circuits coincide with \emph{symplectic transvections}.
For a Hermitian Pauli $P$, the unitary corresponding to a symplectic transvection is given by $\frac{I - \imath P}{\sqrt{2}}$.
When a transvection acts on another Pauli $Q$ by conjugation, it leaves $Q$ unchanged when $P$ and $Q$ commute, and it maps $Q \mapsto \imath Q P$ when they anti-commute.
Hence, symplectic transvections are Clifford operators; in fact, they generate the full Clifford group.
Using this insight, we rigorously show that on a stabilizer code with a logical operator $P$, the transvection defined by $\overline{P}$ yields a physical realization of the transvection defined by $P$ on the logical qubits.
This is done through a systematic analysis of the propagation of stabilizers and $\overline{P}$ on the physical circuit.

Then, we consider the extension to the non-Clifford setting where $\theta \neq \frac{\pi}{2}$.
Here, we establish an important insight: the propagation of a Pauli, Q, on a general Trotter circuit $\exp\left( -\imath \frac{\theta}{2} P \right)$ is a linear combination of itself scaled by $\cos\frac{\theta}{2}$ and the resulting Pauli for $\theta = \frac{\pi}{2}$, i.e., $\exp\left( -\imath \frac{\theta}{2} P \right) Q \exp\left( \imath \frac{\theta}{2} P \right)$, scaled by $\sin\frac{\theta}{2}$.
Using this, we extend the above result for Clifford Trotter circuits (a.k.a. symplectic transvections) to any Trotter circuits with arbitrary $\theta$, i.e., $\exp\left( -\imath \frac{\theta}{2} \overline{P} \right)$ is a physical realization of $\exp\left( -\imath \frac{\theta}{2} P \right)$ for any $[[n,k,d]]$ stabilizer code and any $k$-qubit Pauli operator $P$.
Note that these circuits do not necessarily belong to any level of the Clifford hierarchy, making the result extremely general.
As another insight, we also show that for any Trotter circuit $\exp\left( -\imath \frac{\theta}{2} P \right)$ and any Pauli $Q$ that anti-commutes with $P$, the product of $Q$ and its conjugation by the circuit is $\exp\left( -\imath \theta P \right)$, i.e., a Trotter circuit with double the angle.
We discuss how this helps with examining error propagation.

Finally, we consider the fault tolerance of our approach for a general quantum simulation algorithm that iterates several Trotter circuits and ends with destructive $Z$-measurements.
We propose to use a concatenated QLDPC-cat coding scheme so that the system starts with a heavy bias towards $Z$-errors.
Then, we can leverage bias-preserving gates in cat-based systems to preserve the $Z$-bias through the algorithm.
We schedule syndrome measurements and error correction after each non-trivial Trotter circuit that retains the $Z$-bias.
At the end, the code qubits can also be individually measured in the $Z$-basis, where any residual $Z$-errors have no effect on the output distribution.
This establishes a path towards algorithmic fault tolerance on a single code block in this framework.

\section{Technical Background}
\label{sec:background}

\subsection{Pauli Operators and Stabilizer Codes}
\label{subsec:Pauli_Stabilizer_Code}


For a single qubit, the Hermitian Pauli operators are denoted by \( I \), \( X \), \( Y \), and \( Z \).
For an \( n \)-qubit system, the full Pauli group \( \mathcal{P}_n \) consists of all tensor products of these single-qubit operators, augmented with a global phase of \( \pm 1, \pm \imath \):
\begin{equation}
\mathcal{P}_n = \langle \imath^\kappa I, X, Y, Z \mid \kappa \in \{0, 1, 2, 3\} \rangle^{\otimes n}.
\end{equation}


A stabilizer code is a quantum error correcting code defined by a stabilizer group \( \mathcal{S} \), which is an abelian subgroup of \( \mathcal{P}_n \) that does not include \( -I^{\otimes n} \). The code space is the joint \( +1 \)-eigenspace of all elements in \( \mathcal{S} \), meaning any quantum state $\ket{\psi}$ in the code space satisfies $S\ket{\psi}=\ket{\psi}$, for all $S\in \mathcal{S}$. Logical qubits are encoded within this space, with logical operators, such as \( \overline{X} \) and \( \overline{Z} \), acting non-trivially on the encoded information while commuting with the stabilizers.

Stabilizer codes are typically characterized by the notation $\llbr n,k,d\rrbr$, where $n$ represents the number of physical qubits, $k$ is the number of encoded logical qubits, and $d$ is the code distance, which corresponds to the minimum weight of a Pauli operator that maps one logical codeword to another. The code distance determines the error-correcting capability of the code, allowing it to detect $(d-1)$ and correct up to $\lfloor \frac{d-1}{2} \rfloor$ errors using a maximum-likelihood decoding strategy. 

To represent Hermitian Pauli operators algebraically, a binary formalism is often used. 
We denote such a Pauli as
\begin{equation}
E(a, b) \coloneqq \imath ^{ab^T \bmod\, 4} \bigotimes_{j=1}^n X^{a_j} Z^{b_j},
\end{equation}
where \( a, b \in \mathbb{F}_2^n \) are binary vectors of length \( n \). Here, \( a \) encodes the positions of the \( X \)-components, while \( b \) encodes the positions of the \( Z \)-components. 
For example, for $n=3$, the operator \( X_1 Z_2 X_3 \) can be represented as $E([101, 010])$. 
In this representation, a general Pauli operator acting on \( n \)-qubits maps to a binary vector of length \( 2n \), which can be written as $h= [a \mid b] $, where $h\in\mathbb{F}_2^{2n}$. 
This formalism facilitates efficient algebraic manipulations, such as symplectic transformations, and is particularly valuable in analyzing logical and physical circuits. 
For logical operators, the binary representation can be extended, denoting them with a bar, such as $\overline{h}=\overline{[101 \mid 010]}$. 
Then, \( \overline{E(a, b)}=\overline{E([101,010])} \) represents $ \overline{X_1} \, \overline{Z_2} \,\overline{X_3}$, following the same binary formalism as physical Pauli operators. 
This notation allows for a structured algebraic approach to describe logical operations. 
For instance, in the context of the $\llbr8,3,3\rrbr$ code introduced in Section~\ref{subsec:[[8,3,3]]code}, the logical operator $\overline{X_{1}}$ whose binary representation is $\overline{h_1}=\overline{[100 \mid 000]}$ is physically implemented as $X_{4}X_{5}X_{6}X_{7}$, whose corresponding binary representation is $h_1=[00011110 \mid 00000000]$. 
It is important to note that notation such as $a$, $b$, and $h$ may be context-dependent and, in some cases, overloaded. 
Their precise meaning should always be inferred based on the surrounding discussion, particularly when distinguishing between physical and logical operators within stabilizer codes and fault-tolerant quantum circuits.

\subsection{Symplectic Inner Product and Symplectic Transvections}
\label{subsec:Symplectic_Transvections}

Symplectic transvections play a crucial role in the algebraic structure of stabilizer codes and their associated logical operations. They arise naturally in the binary representation of Pauli operators and are useful for describing the transformation properties of quantum circuits, particularly Clifford circuits. In the binary formalism, Pauli operators are represented as length-$2n$ binary vectors, forming a $2n$-dimensional symplectic vector space over $\mathbb{F}_2$.

The symplectic structure of this space is defined through the symplectic inner product (Def.~\ref{def:symplectic_inner_product}), which provides a means of determining commutation and anti-commutation relations between Pauli operators. 
\begin{definition}
\label{def:symplectic_inner_product}
The symplectic inner product between two Pauli operators, represented in binary form as \( a = [a_1 \mid a_2] \) and \( b = [b_1 \mid b_2] \), where $a_1,a_2,b_1,b_2\in\mathbb{F}_2^n$, is given by:
\begin{equation}
\label{eq:symplectic_inner_product}
\langle a, b \rangle_s = a_1 \cdot b_2^T + a_2 \cdot b_1^T (\bmod\ 2) = a\, \Omega\, b^T,
\end{equation}
where $\Omega=\begin{bmatrix}
    0 & -I_n\\
    I_n & 0\\
\end{bmatrix} =\begin{bmatrix}
    0 & I_n\\
    I_n & 0\\
\end{bmatrix}$.
\end{definition}
This quantity determines whether two Pauli operators commute (\(\langle a, b \rangle_s = 0\)) or anti-commute (\(\langle a, b \rangle_s = 1\)).    

A symplectic transvection (Def.~\ref{def:symplectic transvection}) is a linear transformation that preserves this inner product while applying a specific transformation to a given Pauli operator. 

\begin{definition}
\label{def:symplectic transvection}
Given a vector $h = [a \mid b] \in \mathbb{F}_2^{2n}$, a \emph{symplectic transvection} is a map $Z_h : \mathbb{F}_2^{2n} \to \mathbb{F}_2^{2n}$ defined by
\begin{align}
Z_h(x) = x + \langle x, h \rangle_{s} h \quad \iff \quad F_h = I_{2n} + \Omega h^T h,
\end{align}
where $F_h$ represents its associated symplectic matrix.

The unitary operator corresponding to the symplectic transvection $h$ is given by $U_h(\frac{\pi}{2})$, where
\begin{equation} 
\label{eq:Unitary_operator_h}
U_h(\theta) := \exp\left(-\imath \frac{\theta}{2} E(a,b) \right) 
\end{equation} 
is a \emph{generalized symplectic transvection}.
\end{definition}

Symplectic transvections are Clifford operators and are known to generate the full Clifford group.
Hence, they form an alternative set of generators besides the standard CNOT, Hadamard, and Phase gates.
There exists a systematic algorithm to identify a sequence of at most $2t$ transvections that together satisfy a valid set of $t$ linear equations~\cite{Koenig-jmp14,Rengaswamy-tqe20}.


\subsection{$\llbr 8,3,3\rrbr$ non-CSS Code} 
\label{subsec:[[8,3,3]]code}

A key example of a stabilizer code used throughout this work is the \(\llbr 8,3,3 \rrbr\) code~\cite{Chao-prl18}, which is a non-CSS stabilizer code. 
In general, a Calderbank-Shor-Steane (CSS) code is a subclass of stabilizer codes where the stabilizer group can be generated by elements consisting entirely of either \(X\)-type or \(Z\)-type Pauli operators. This structure enables transversal implementation of certain logical gates, making CSS codes particularly useful in fault-tolerant quantum computation.

However, the \(\llbr 8,3,3 \rrbr\) code does not follow this form; its stabilizers contain mixed \(X\)- and \(Z\)-type terms. 
Despite this, the techniques and results presented in this paper are apply to even such codes.
The logical operators and stabilizers of the \(\llbr 8,3,3 \rrbr\) code are given as follows:
\begin{center}
\begin{tabular}{|l|l|}
\toprule
 $S_{1} = X_{1}X_{2}X_{3}X_{4}X_{5}X_{6}X_{7}X_{8}$ & $\overline{X_1} = {X_{4}X_{5}X_{7}X_{8}}$ \\
 $S_{2} = Z_{1}Z_{2}Z_{3}Z_{4}Z_{5}Z_{6}Z_{7}Z_{8}$ & $\overline{X_2} = {X_{3}Z_{4}Z_{5}X_{6}}$ \\
 $S_{3} = Z_{3}Y_{4}X_{5}Z_{6}Y_{7}X_{8}$ & $\overline{X_3} = {Z_{1}Z_{2}X_{6}X_{7}}$ \\
 $S_{4} = Z_{2}X_{3}X_{5}Y_{6}Z_{7}Y_{8}$ & $\overline{Z_1} = {Z_{2}X_{3}Z_{5}X_{8}}$ \\ 
 $S_{5} = X_{2}Z_{4}Z_{5}X_{6}Y_{7}Y_{8}$ & $\overline{Z_2} = {Z_{1}Z_{5}Z_{6}Z_{7}}$ \\
 (This is a non-degenerate code)  & $\overline{Z_3} = {Z_{1}Z_{2}Z_{4}Z_{7}}$ \\
\bottomrule
\end{tabular}
\end{center}


This formulation of the \(\llbr 8,3,3 \rrbr\) code provides a concrete example for demonstrating the techniques developed in this work. Nevertheless, the results extend beyond this specific case and hold for any stabilizer code, including those with different structures and distance properties, such as quantum low-density parity-check (QLDPC) codes.

\subsection{Quantum Hamiltonian Simulation and Trotter circuits}
\label{subsec:Trotter_circuits}

In quantum Hamiltonian simulation, the goal is to approximate the evolution of a quantum state under a Hamiltonian \( H \) for a given time \( t \). The Hamiltonian is typically expressed in terms of Hermitian Pauli operators \( E_j \), as in~\cite{Whitfield-molphy11}:
\begin{equation}
\label{eq:sum_Hamiltonian}
    H = \sum_j \alpha_j E_j, \quad \alpha_j \in \mathbb{R}.
\end{equation}

The time evolution operator is given by the unitary operator:
\begin{equation}
\label{eq:time_revolution}
    U(t) = \exp(-\imath Ht) = \exp\left(-\imath \sum_j \alpha_j E_j t\right),
\end{equation}
which can be decomposed into a sequence of exponentials for each \( E_j \) using the Trotter-Suzuki approximation~\cite{Hatano-qaoom05}:
\begin{equation}
\label{eq:Trotter_Suzuki_approximation}
    U(t) \approx \left( \prod_j \exp\left(-\imath \frac{\alpha_j E_j t}{T}\right) \right)^T,
\end{equation}
where \( T \) is the number of Trotter steps.

In this context, the circuit implementing \( U_h(\theta) \) for a single Pauli term \( E(a,b) \) is called a Trotter circuit or a quantum simulation kernel (QSK) circuit. Here, \( \theta = 2\alpha_j t/T \) represents the rotation angle. The Trotter circuit exhibits a highly symmetric structure, characterized by:
\begin{itemize}
    \item The balanced arrangement of Hadamard and \( H_y \)($H_y \coloneqq \frac{1}{\sqrt{2}} \begin{bsmallmatrix}
1 & -\imath \\ \imath & -1 \end{bsmallmatrix}$) gates at the start and end of the circuit.
    \item A mirrored sequence of CNOT gates applied symmetrically around the central qubit.
    \item The central \( R_z(\theta) = \exp\left( -\imath \frac{\theta}{2} Z \right) \) gate
acting on the highest index qubit within the support of \( E(a,b) \).
\end{itemize}
An example is shown later in Fig.~\ref{fig:3_qubit_non_Clifford_logical}.
%
For a Trotter circuit acting on \( n \)-qubits, the circuit depth is linear in the weight \( w \) of \( E(a,b) \), specifically \( 2w + 1 \), where \( w \) is the number of qubits on which \( E(a,b) \) acts non-trivially. 
This symmetry not only ensures efficient implementation but also simplifies the analysis of operator propagation through the circuit.


\section{Realization of Clifford Trotter circuit on any Stabilizer Code}
\label{sec:clifford_trotter}

Trotter circuits provide a framework for decomposing unitary operations. When implemented on stabilizer codes, they enable reliable quantum computation while manifesting a clear correspondence between logical operations and physical implementations thereof. This correspondence is essential for fault-tolerant quantum computation, as it ensures that the structural symmetry and consistency of the Trotter circuit are preserved across logical and physical layers. A key insight underlying this realization is that logical Trotter circuits and their physical implementations share the same Trotter pattern, regardless of the stabilizer code used. This symmetry enables seamless translation between logical and physical circuits, preserving stabilizer properties and structural consistency.

To establish this connection, we analyze the propagation of logical Pauli operators through Clifford Trotter circuits and introduce the framework of symplectic transvections, which play a central role in maintaining the Trotter pattern during logical-to-physical mapping. The propagation of logical Pauli operators in Trotter circuits follows a systematic mapping that ensures consistency throughout the circuit. This Pauli mapping plays a vital role in maintaining the symmetry of Trotter circuits and supporting fault-tolerant quantum computation on stabilizer codes. To formalize this mapping, we introduce notation to describe the logical qubits and their transformations under Clifford Trotter circuits.

Let $[k] \coloneqq \{1, 2, 3, \dots, k\}$ denote the index set of logical qubits in a Clifford Trotter circuit, where $k$ represents the number of logical qubits in an \(\llbr n, k, d\rrbr\) stabilizer code. Define $I_h \subseteq [k]$ as the subset of indices where Hadamard gates are applied at the beginning and the end of the Trotter circuits. Similarly, let $I_{hy} \subseteq [k]$ represent the subset of indices where $H_y$ gates are applied. Logical qubits that do not contain either Hadamard or $H_y$ gates belong to the subset $I_e \coloneqq [k] \backslash (I_h \cup I_{hy})$. Importantly, $I_h$ and $I_{hy}$ are disjoint. Using this notation, the propagation rules for logical Pauli operators in Clifford Trotter circuits are described next.

\begin{lemma}
\label{lem:c_qsk_logical_Pauli_mappings}
Logical Pauli mappings of the general Clifford Trotter circuit on $k$ qubits can be expressed as follows:
\begin{enumerate}

\item If $i \notin I_{h} \cup I_{hy}$, or $i \in I_{e}$, then
\begin{align}
\overline{X_{i}} & \mapsto \left( \prod_{j\in I_{h}} \overline{X_{j}} \right) \left( \prod_{j\in I_{hy} \cup \{i\}} \overline{Y_{j}} \right)\left( \prod_{j \in I_{e} \setminus \{ i \}} \overline{Z_{j}} \right), \\
\overline{Z_{i}} & \mapsto \overline{Z_{i}}.
\end{align}

\item If $i \in I_{h}$, then
\begin{align}
\overline{X_{i}} & \mapsto \overline{X_{i}}, \\    
\overline{Z_{i}} &\mapsto - \left( \prod_{j \in I_{h} \setminus \{ i \}}\overline{X_{j}} \right) \left( \prod_{j \in I_{hy} \cup \{ i \}}\overline{Y_{j}} \right) \left(\prod_{j \in I_{e}} \overline{Z_{j}} \right).    
\end{align}

\item If $i \in I_{hy}$, then
\begin{align}
\overline{X_{i}} & \mapsto -\left( \prod_{j\in I_{h}} \overline{X_{j}} \right) \left( \prod_{j\in I_{hy} \setminus \{i\}} \overline{Y_{j}} \right)\left( \prod_{j \in I_{e} \cup \{ i \}} \overline{Z_{j}} \right), \\
\overline{Z_{i}} &\mapsto \left( \prod_{j \in I_{h} \cup \{i\}}\overline{X_{j}} \right) \left( \prod_{j \in I_{hy} \setminus \{ i \}}\overline{Y_{j}} \right) \left(\prod_{j \in I_{e}} \overline{Z_{j}} \right).
\end{align}
\end{enumerate}
\end{lemma}

This lemma provides a complete characterization of how logical operators $\overline{X}$ and $\overline{Z}$ propagate through arbitrary logical Clifford Trotter circuits. 
The detailed proof of Lemma~\ref{lem:c_qsk_logical_Pauli_mappings} is omitted here for brevity. 
The proof follows a similar methodology as in our previous work~\cite{Chen-qce24,Chen-arxiv24}, where the propagation of logical Pauli operators under Clifford Trotter circuits without $H_y$ gates was rigorously established. 
The inclusion of $H_y$ gates requires only minor modifications to the original proof, as the underlying approach remains consistent. 

To understand how these logical transformations manifest in physical implementations, it is useful to employ a framework that captures the relationship between logical and physical operations. 
One such tool is the symplectic transvection (Def.~\ref{def:symplectic transvection}), which provides a systematic way to map logical operators to their physical counterparts while preserving the stabilizer properties and the structural symmetry of Trotter circuits.



\begin{theorem}
\label{thm:Clifford any stabilizer code}
For an arbitrary logical Clifford circuit on any \(\llbr n, k, d \rrbr\) stabilizer code, let \(\bar{h} \in \mathbb{F}_2^{2n} \) denote a symplectic transvection associated with this logical circuit. 
Consider two conditions on how the circuit conjugates input Pauli operators to output Pauli operators:
\begin{enumerate}
    \item For every non-trivial Pauli constraint, the input and output Pauli operators anti-commute, and their binary representations sum to \(\bar{h}\);

    \item For every trivial Pauli constraint whose input and output Pauli operators are the same, the operators therein commute with \( E(\bar{h}) \).
\end{enumerate}
If these are satisfied for all input Pauli constraints of the circuit, then the symplectic transvection \(\bar{h}\) implements this logical Clifford circuit and commutes with all stabilizers.
\begin{IEEEproof}
\normalfont
Define $h \coloneqq x \oplus y$, where $x, y \in \mathbb{F}_2^{2n}$ are binary representations of input and output Pauli operators for a Pauli constraint of the physical circuit, respectively.
\begin{enumerate}
\item 
For non-trivial Pauli constraints, if $x$ and $y$ anti-commute, i.e., $\langle x,y \rangle_{s}=1$, then
\begin{align}
\label{eq:non-trivial-begin}
    Z_{\bar{h}}(x) & = x F_{\bar{h}} \\ 
                   & = x+\langle x,\bar{h} \rangle_{s} \bar{h} \\
                   & = x+\langle x, x+y \rangle_{s} \bar{h} \\
                   & = x+ \langle x,x \rangle_{s} \bar{h} + \langle x,y \rangle_{s} \bar{h} \\
                   & = x+\bar{h} \\
                   & = x+x+y \\
                   & = y.
\label{eq:non-trivial-end}
\end{align}

\item 
For trivial Pauli constraints, including stabilizers, $x=y$, and $\langle x,\bar{h} \rangle_{s}=0$, so we have
\begin{align}
\label{eq:trivial}
    Z_{\bar{h}}(x) = x F_{\bar{h}} = x + \langle x,\bar{h} \rangle_{s} \bar{h} = x.
\end{align}
\end{enumerate}

Hence, for the non-trivial Pauli constraints, the input Pauli $x$ always maps to the correct output Pauli $y$ (Eqns.~\eqref{eq:non-trivial-begin}-\eqref{eq:non-trivial-end}), and symplectic transvection never operates on the trivial Pauli constraints (Eqn.~\eqref{eq:trivial}). 
Therefore, $F_{\bar{h}}$ implements the logical circuit and stabilizers are preserved.
\end{IEEEproof}
\end{theorem}

Logical Trotter circuits satisfy the conditions of Theorem~\ref{thm:Clifford any stabilizer code}. 
Specifically, the symmetric arrangement of gates in Trotter circuits ensures that the propagation of Pauli operators adheres to the commutation and anti-commutation relationships required by Theorem~\ref{thm:Clifford any stabilizer code}. This is illustrated in Corollary~\ref{cor:C-QSK centralizes stabilizer}.

\begin{corollary}
\label{cor:C-QSK centralizes stabilizer}
On any $\llbr n,k,d \rrbr$ stabilizer code, the transvection $\bar{h}$ implements a logical Clifford Trotter circuit and preserves all stabilizers, i.e, centralizes the stabilizer group.
\begin{IEEEproof}
\normalfont
From Lemma~\ref{lem:c_qsk_logical_Pauli_mappings}, multiplying non-trivial input and output Pauli strings of the logical Clifford Trotter circuit gives:
\begin{align}
\label{eq:transvection of C-QSK}
E(\bar{h}) = \imath\left( \prod_{i\in I_{h}} \overline{X_{i}} \right) \left( \prod_{j\in I_{hy}} \overline{Y_{j}} \right)\left( \prod_{k \in I_{e}} \overline{Z_{k}} \right).
\end{align}
This transvection \(\bar{h}\) representing $U_{\bar{h}}(\frac{\pi}{2})$ describes the overall transformation of the Clifford Trotter circuit. To analyze its properties, we consider whether the input and output Pauli operators commute or anti-commute, as required by Lemma~\ref{lem:c_qsk_logical_Pauli_mappings}. 

For the first scenario in Lemma~\ref{lem:c_qsk_logical_Pauli_mappings}, the mapping of \(\overline{X_i}\) is:
\begin{align}
\label{eq:1_non_trivial}
\overline{X_i} \mapsto \left( \prod_{j \in I_h} \overline{X_j} \right) \left( \prod_{j \in I_{hy} \cup \{i\}} \overline{Y_j} \right) \left( \prod_{j \in I_e \setminus \{i\}} \overline{Z_j} \right).
\end{align}
To verify the anti-commutation property for a non-trivial Pauli constraint, we examine the terms in the output involving the index \(i\). The term \(\overline{Y_i}\) appears in (\(\prod_{j \in I_{hy} \cup \{i\}} \overline{Y_j}\)), ensuring that the input \(\overline{X_i}\) and the output anti-commute due to the Pauli relation \(\overline{X_i} \, \overline{Y_i} = -\overline{Y_i} \, \overline{X_i}\). 
For the trivial Pauli constraint in this scenario, where \(\overline{Z_i} \mapsto \overline{Z_i}\), the operator remains unchanged, and since \(i \in I_e\), it commutes with $U_{\bar{h}}$. 

The same reasoning applies to the remaining two scenarios in Lemma~\ref{lem:c_qsk_logical_Pauli_mappings}. In both cases, the anti-commutation or commutation of input and output operators is determined by isolating the terms associated with the relevant index \(i\).

Therefore, the transvection \(\bar{h}\) of the logical Clifford Trotter circuit in Eqn.~(\ref{eq:transvection of C-QSK}) satisfies the hypotheses in Theorem~\ref{thm:Clifford any stabilizer code} and thus implements a logical Clifford Trotter circuit and preserves all stabilizers, thereby centralizing the stabilizer group.
\end{IEEEproof}
\end{corollary}

An important characteristic of Trotter circuits is that their logical and physical implementations share the Trotter pattern. This shared symmetry arises naturally from the structure of the symplectic transvection, which is expressed in Eqn.~(\ref{eq:transvection of C-QSK}). 

\begin{remark}
\label{rem:physical_clifford_Trotter}
For any logical Clifford Trotter circuit on an arbitrary stabilizer code, substituting the stabilizer code's logical operators (e.g., $\overline{X}$, $\overline{Y}$, $\overline{Z}$) into Equation~\ref{eq:transvection of C-QSK} results in a physical circuit that maintains a Trotter pattern on the qubits involved (referred to as the support of the logical operators). In the physical circuit, symmetric CNOT gates are introduced with control qubits on all these qubits except for the one with the highest index, which serves as the target qubit. A phase gate at this highest index qubit in the middle separates the CNOT gates symmetrically. For the symplectic transvection components, qubits corresponding to the $X$-components are added with Hadamard gates, while those corresponding to the $Y$-components are added with $H_y$ gates, placed symmetrically on the left and right sides of the physical circuit. 
\end{remark}

To illustrate the construction of physical Trotter circuits, we now present an example using the $\llbr 8,3,3 \rrbr$ non-CSS stabilizer code. 
This example demonstrates how the symplectic transvection derived from the logical circuit can be translated into a physical implementation, preserving the Trotter pattern and ensuring stabilizer centralization. 

\begin{figure}[h]
\centering
\includegraphics[scale=1,keepaspectratio]{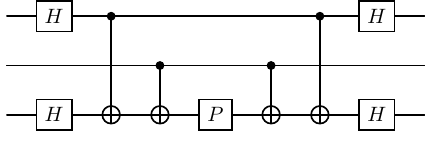}

\caption{Logical Clifford Trotter circuit on $k=3$ qubits. Here, $P$ refers to the Clifford Phase gate $R_Z(\frac{\pi}{2})$.}
\label{fig:qsk_clifford_3qubit}

\end{figure}

The mappings for the logical Trotter circuit in Fig.~\ref{fig:qsk_clifford_3qubit} are:
\begin{IEEEeqnarray}{lClCrCl}
\label{eq:logical_qsk_3qubit_constraints}
\overline{X_{1}} \mapsto \overline{X_{1}} & \quad , \quad & \overline{Z_{1}} \mapsto -\overline{Y_{1}} \, \overline{Z_{2}} \, \overline{X_{3}} \ , \nonumber \\
\overline{X_{2}} \mapsto \overline{X_{1}} \, \overline{Y_{2}} \,\overline{X_{3}} & \quad , \quad & \overline{Z_{2}} \mapsto \overline{Z_{2}} \ , \nonumber \\
\overline{X_{3}} \mapsto \overline{X_{3}} & \quad , \quad & \overline{Z_{3}} \mapsto -\overline{X_{1}} \, \overline{Z_{2}} \, \overline{Y_{3}} \ .
\end{IEEEeqnarray}
For all non-trivial Pauli constraints in Eqn.~(\ref{eq:logical_qsk_3qubit_constraints}), multiplying their Pauli inputs and outputs yields:
\begin{align}
\label{eq:symplectic_transvection_multipication}
\overline{X_{2}}\cdot(\overline{X_{1}} \, \overline{Y_{2}} \,\overline{X_{3}})=\imath\overline{X_{1}} \, \overline{Z_{2}} \, \overline{X_{3}}, \\
\overline{Z_{1}}\cdot(-\overline{Y_{1}} \, \overline{Z_{2}} \, \overline{X_{3}})=\imath\overline{X_{1}} \, \overline{Z_{2}} \, \overline{X_{3}}, \\
\overline{Z_{3}}\cdot(-\overline{X_{1}} \, \overline{Z_{2}} \, \overline{Y_{3}})=\imath\overline{X_{1}} \, \overline{Z_{2}} \, \overline{X_{3}}. 
\end{align}
As shown, the symplectic transvection derived from the multiplication of all non-trivial Pauli constraints is $E(\bar{h}) = \imath \overline{X_{1}} \, \overline{Z_{2}} \, \overline{X_{3}}$.
This demonstrates that the symplectic transvection is fixed and is consistent with the one derived in Corollary~\ref{cor:C-QSK centralizes stabilizer}. Moreover, for non-trivial Pauli constraints, the input and output Pauli operators always anti-commute. In contrast, trivial Pauli constraints commute with the symplectic transvection.

To establish a direct connection between the logical Trotter circuit and its physical implementation, we embed the logical circuit in Fig.~\ref{fig:qsk_clifford_3qubit} within the $\llbr 8,3,3 \rrbr$ non-CSS code. 
By substituting the logical operators of the $\llbr 8,3,3 \rrbr$ code into the mappings for the logical Trotter circuit in Fig.~\ref{fig:qsk_clifford_3qubit}, the following physical Pauli constraints are obtained:
\begin{align}
\label{eq:8_qubit_constraints_physical}
X_{4}X_{5}X_{7}X_{8} & \mapsto  X_{4}X_{5}X_{7}X_{8}, \nonumber \\
X_{3}Z_{4}Z_{5}X_{6} & \mapsto  Z_{2}X_{3}Y_{4}X_{5}Z_{6}Z_{7}X_{8}, \nonumber \\ 
Z_{1}Z_{2}X_{6}X_{7} & \mapsto  Z_{1}Z_{2}X_{6}X_{7}, \nonumber \\
Z_{2}X_{3}Z_{5}X_{8} & \mapsto  X_{3}X_{4}X_{5}Y_{6}Z_{7}, \nonumber \\ 
Z_{1}Z_{5}Z_{6}Z_{7} & \mapsto  Z_{1}Z_{5}Z_{6}Z_{7}, \nonumber \\
Z_{1}Z_{2}Z_{4}Z_{7} & \mapsto  Z_{1}Y_{4}Y_{5}Y_{6}X_{8}.
\end{align}



\begin{figure}[t]
    \centering
    \includegraphics[width=1\linewidth]{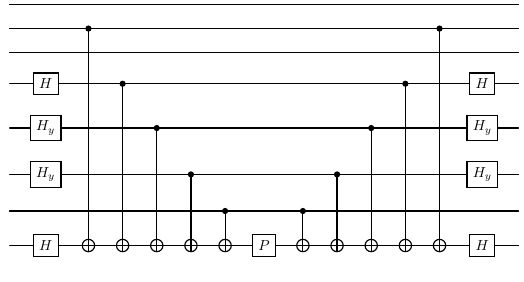}
    \caption{Physical realization of Fig.~\ref{fig:qsk_clifford_3qubit} on $\llbr 8,3,3 \rrbr$ code. Here, $P$ refers to the Clifford Phase gate $R_Z(\frac{\pi}{2})$.}
    \label{fig:8_physical_Clifford_Trotter}
\end{figure}

As described in Remark~\ref{rem:physical_clifford_Trotter}, the symplectic transvection result, \( \overline{X_{1}} \, \overline{Z_{2}} \, \overline{X_{3}} = Z_2 X_4 Y_5 Y_6 Z_7 X_8 \), guides the construction of the physical circuit. 
The full physical circuit is displayed in Fig.~\ref{fig:8_physical_Clifford_Trotter}. 
We can verify that the physical circuit matches the logical Pauli constraints derived earlier in Eqn.~(\ref{eq:8_qubit_constraints_physical}), thus implementing the mappings. 
Additionally, the stabilizer group of the $\llbr 8,3,3 \rrbr$ code is centralized by the circuit.




We have not yet considered the role of stabilizers in the context of symplectic transvection. It is well-known that for any logical operator, multiplying it by a stabilizer does not change its logical effect. This property also holds for the symplectic transvection. By carefully selecting stabilizers and multiplying them with the symplectic transvection, we can reduce the support of the symplectic transvection, thereby further decreasing the depth of the physical circuit.

\begin{figure}
    \centering
    \includegraphics[width=0.8\linewidth]{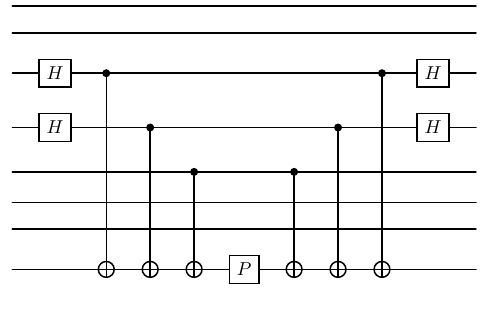}
    \caption{The updated physical circuit after considering the effect of a stabilizer to reduce the depth. By multiplying the original symplectic transvection with the stabilizer \(S_4\), the support size of the symplectic transvection is reduced, leading to fewer CNOT gates and an overall reduction in the circuit depth.}
    \label{fig:8_qubit_after_stabilizer}
\end{figure}

For the previously discussed $\llbr 8,3,3 \rrbr$ code example, let us take the original symplectic transvection $Z_2X_4Y_5Y_6Z_7X_8$ and multiply it by the stabilizer $S_{4} = Z_{2}X_{3}X_{5}Y_{6}Z_{7}Y_{8}$. The resulting new symplectic transvection becomes $X_{3}X_{4}Z_{5}Z_{8}$ (up to a sign). 
The support size of the symplectic transvection is now reduced to 4, which further reduces the number of required CNOT gates in the physical circuit. 
The new physical circuit is shown in Fig.~\ref{fig:8_qubit_after_stabilizer}.

The analysis of the $\llbr8,3,3\rrbr$ stabilizer code demonstrates a broader principle. For any logical Clifford Trotter circuit implemented on any stabilizer code, the corresponding symplectic transvection retains its fundamental properties. These include the ability to centralize stabilizers, preserve the Trotter pattern, and allow for circuit depth optimization by carefully selecting and incorporating stabilizers. These properties ensure a consistent logical-to-physical mapping and underscore the versatility of symplectic transvection as a framework for fault-tolerant quantum computation.

While this work primarily focuses on the Clifford case, it is worth noting that similar principles apply in the presence of non-Clifford gates. The inclusion of gates such as $R_z(\theta)$ introduces additional considerations, including the potential for more complex Pauli propagation. However, the fundamental concepts of stabilizer centralization and symplectic transvection remain relevant, providing a foundation for extending these techniques to non-Clifford circuits.


\section{Extension to Non-Clifford Trotter Circuits on Stabilizer Codes}
\label{sec:non_clifford_trotter}


To generalize our framework from the Clifford case to non-Clifford Trotter circuits, we present a series of insights that collectively extend the properties of symplectic transvections to arbitrary-angle $R_z(\theta)$ gates in logical Trotter circuits. 
These results address three key aspects: the propagation of logical operators (Theorem~\ref{thm:non_Clifford_logical_operators_propagation}), the preservation and centralization of stabilizers (Theorem~\ref{thm:non_Clifford_stabilizers}), and the compatibility of non-Clifford Trotter circuits with a generalized transvection framework (Theorem~\ref{thm:non_Clifford_product}). 
Together, they establish a robust theoretical foundation for implementing logical non-Clifford Trotter circuits while maintaining symmetry and stabilizer-centralization.

To illustrate Theorem~\ref{thm:non_Clifford_logical_operators_propagation} and Theorem~\ref{thm:non_Clifford_stabilizers}, we return to the $\llbr8,3,3\rrbr$ stabilizer code. For each theorem, we demonstrate its implications through explicit constructions and Pauli propagation mappings in physical circuits. While Theorem~\ref{thm:non_Clifford_product} does not involve a specific example, it unifies the framework by formally extending transvection beyond the symplectic regime, enabling a consistent analysis of logical and physical mappings in non-Clifford Trotter circuits. By doing so, it highlights the compatibility of non-Clifford gates with the broader structure of Trotter circuits and ensures that the foundational principles established in the Clifford case remain applicable.

We now formalize the propagation of logical operators in non-Clifford Trotter circuits with the following theorem:
\begin{theorem}
\label{thm:non_Clifford_logical_operators_propagation}
Consider a logical non-Clifford Trotter circuit, i.e., $U_h(\theta)$ with an arbitrary rotation angle, on any stabilizer code. 
Construct a physical Trotter circuit by first synthesizing the corresponding logical Clifford Trotter circuit, $U_{\bar{h}}(\frac{\pi}{2})$, and then replacing the Phase gate with $R_z(\theta)$. 
Then this physical circuit satisfies all Pauli constraints on the logical circuit.
\begin{IEEEproof}
\normalfont
To analyze the propagation of a general logical Pauli operator through the central \( R_z(\theta) \) gate in the non-Clifford logical Trotter circuit, we start with the conjugation rules for \( R_z(\theta) \) acting on the Pauli operators \( X \) and \( Y \):
\begin{align}
R_z(\theta) X R_z^\dagger(\theta) &= \cos(\theta) X + \sin(\theta) Y \\
                                  &= \cos(\theta) X + \sin(\theta)(P X P^\dagger), \\
\label{eq:Rz_X}
R_z(\theta) Y R_z^\dagger(\theta) &= \cos(\theta) Y - \sin(\theta) X \\
                                  &= \cos(\theta) Y + \sin(\theta)(P Y P^\dagger),
\label{eq:Rz_Y}
\end{align}
where \( P = R_Z(\frac{\pi}{2}) \) denotes the Phase gate acting on the highest index qubit within the support of the symplectic transvection in the Clifford case. 
Since $P X P^\dagger=Y$ and $P Y P^\dagger=-X$, these equations show that in the non-Clifford case, $R_z(\theta)$ decomposes the conjugation of $X$ or $Y$ operators into two terms: 
the original input operator itself scaled by $\cos(\theta)$, and 
a Clifford-like contribution scaled by $\sin(\theta)$, corresponding to conjugation by $P$.
For \( Z \) and \( I \), the conjugation map acts trivially: $R_z(\theta) Z R_z^\dagger(\theta) = Z, R_z(\theta) I R_z^\dagger(\theta) = I$.
Thus, for Pauli operators other than $X$ or $Y$, the action of $R_z(\theta)$ is identical to that of the Clifford case, regardless of the rotation angle.
Due to the linearity of conjugation, which can be expressed as $A (B + C) A^\dagger = A B A^\dagger + A C A^\dagger$, we may treat the two contributions separately. After passing through $R_z(\theta)$, the propagation continues through the remaining Clifford gates in the circuit, including a series of CNOT gates and Hadamard ($H$) or $H_y$ gates at the end. 
Crucially, the $R_z(\theta)$ gate affects only the last qubit, ensuring that the propagation of operators on other qubits remains consistent with the Clifford case.

In the physical Trotter circuit, for the non-trivial Pauli constraints, the propagation of a general Pauli operator \( E \) through the central \( R_Z(\theta) \) gate can be expressed as follows. 
Let \( E \) be a general operator of the form:
    $E = E_{\text{rest}} \otimes E_{\text{highest}}$,
where \( E_{\text{rest}} \) acts on all qubits except the highest index within the support of the symplectic transvection (defined in the corresponding Clifford Trotter circuit), and \( E_{\text{highest}} \) acts solely on the highest index qubit. 
Then the propagation through \( R_Z(\theta) \) can now be written as:
\begin{equation}
\label{eq:physical_propagation}
    R_z(\theta) E R_z^\dagger(\theta) = E_{\text{rest}} \otimes R_z(\theta) E_{\text{highest}} R_Z^\dagger(\theta).
\end{equation}
Substituting the conjugation formulas for \( E_{\text{highest}} \), we get:
\begin{align}
\label{eq:physical_propagation_final}
    R_z(\theta) E R_z^\dagger(\theta) &= \cos(\theta) \big(E_{\text{rest}} \otimes E_{\text{highest}} \big) \nonumber\\ & \qquad + \sin(\theta) \big(E_{\text{rest}} \otimes P E_{\text{highest}} P^\dagger \big) \nonumber\\
    &=\cos(\theta) E + \sin(\theta) \big(E_{\text{rest}} \otimes P E_{\text{highest}} P^\dagger \big),
\end{align}
where the contributions are split into two terms. 
The first term corresponds to $E$ itself, scaled by $\cos(\theta)$, while the second term involves a Clifford-like contribution, scaled by $\sin(\theta)$. 
The linearity of conjugation ensures that these contributions can be treated independently and combined at the end, ensuring that non-trivial Pauli constraints are still satisfied by the non-Clifford physical Trotter circuit.
For trivial Pauli constraints, the propagation remains unaffected by the central $R_z(\theta)$ gate, akin to the behavior in the Clifford case. 
Hence, $U_{\bar{h}}(\theta)$ satisfies all Pauli constraints of $U_h(\theta)$ for any $\theta$.
\end{IEEEproof}    
\end{theorem}

\begin{remark}
    An alternative way to see Theorem \ref{thm:non_Clifford_logical_operators_propagation} is as follows:
    Promote $\overline{\phantom{A}}: U(2^k) \to U(2^n)$ by $\mathbb{C}$-linearity to a homomorphism of group algebras $\mathbb{C}[U(2^k)] \to \mathbb{C}[U(2^n)]$, i.e.,
    \begin{align}
        \overline{A B} &= \overline{A} \phantom{i} \overline{B},
        \overline{a A + b B} &= a \overline{A} + b \overline{B},
    \end{align}
    where $a, b \in \mathbb{C}$ and $A, B \in \mathbb{C}[U(2^k)]$.
    For $A \in \mathbb{C}[U(2^k)]$, 
    \begin{equation}
        \overline{\exp(\imath \theta A)} = \exp (\overline{\imath \theta A}).
    \end{equation}
    To see this, observe that
    \begin{multline}
        \overline{\exp(\imath \theta A)} = \overline{\sum_{m=0}^\infty \frac{\imath \theta A^m}{m!}} = \sum_{m=0}^\infty \frac{\overline{\imath \theta A^m}}{m!} = \\
        \sum_{m=0}^\infty \frac{\overline{\imath \theta A}^m}{m!} = \exp(\overline{\imath \theta A}).
    \end{multline}
\end{remark}


\begin{remark}
\label{rem:non_Clifford_propagation_contributions}
    For an arbitrary non-Clifford Trotter circuit, the input-output relationship for non-trivial Pauli constraints is:
    \begin{equation}
    \label{eq:physical_Pauli_contributions}
    U_h(\theta) E_{\text{in}} U_h^\dagger(\theta) = \cos(\theta) E_{\text{in}} + \sin(\theta) U_h\left(\frac{\pi}{2}\right) E_{\text{in}} U_h^\dagger\left(\frac{\pi}{2}\right),
    \end{equation}
    where \( U_h(\theta) \) represents the unitary operator of the non-Clifford Trotter circuit with an arbitrary \( Z \)-rotation angle, as defined in Eqn.~(\ref{eq:Unitary_operator_h}), and \( E_{\text{in}} \) denotes the input Pauli operator. 

    For the corresponding logical Trotter circuit, the relationship for logical operators is analogous:
    \begin{equation}
    \label{eq:logical_Pauli_contributions}
    U_{\overline{h}}(\theta) \overline{E_{\text{in}}} U_{\overline{h}}^\dagger(\theta) = \cos(\theta) \overline{E_{\text{in}}} + \sin(\theta) U_{\overline{h}}\left(\frac{\pi}{2}\right) \overline{E_{\text{in}}} U_{\overline{h}}^\dagger\left(\frac{\pi}{2}\right),
    \end{equation}
    where \( U_{\overline{h}}(\theta) \) is the logical unitary operator, and \( \overline{E_{\text{in}}} \) represents the logical Pauli operator corresponding to \( E_{\text{in}} \).
\end{remark}


    

With the proof of the Theorem~\ref{thm:non_Clifford_logical_operators_propagation} and the summary of the Remark~\ref{rem:non_Clifford_propagation_contributions} complete, we now turn to an explicit example using the $\llbr 8,3,3 \rrbr$ code we used before to demonstrate how non-trivial Pauli constraints are satisfied in the physical Trotter circuit. This example will illustrate the propagation characteristics derived earlier and validate that the mappings of logical operators remain consistent even when the central $R_z(\theta)$ gate is introduced. Specifically, we will show how the decomposition of Pauli operators, as outlined in the proof, ensures the preservation of the logical structure while satisfying all non-trivial Pauli constraints in the physical circuit.

\begin{figure}[t]
    \centering
    \includegraphics[scale=1,keepaspectratio]{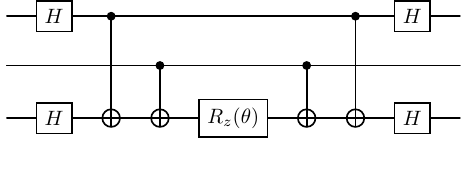}
    \caption{Non-Clifford logical Trotter circuit with 3 qubits, where the central Phase gate in Fig.~\ref{fig:qsk_clifford_3qubit} is replaced by $R_z(\theta)$.}
    \label{fig:3_qubit_non_Clifford_logical}
\end{figure}

To illustrate the behavior of a non-Clifford Trotter circuit, we consider the 3-qubit logical Trotter circuit, which is modified from its Clifford counterpart by replacing the central Phase gate with a general $R_z(\theta)$ gate of arbitrary angle. 
The mappings for non-Clifford logical Trotter circuit in Fig.~\ref{fig:3_qubit_non_Clifford_logical} are:
\begin{align}
\label{eq:logical_non_Clifford_qsk_3qubit_constraints}
\overline{X_{1}} \mapsto \overline{X_{1}} \ , \ 
\overline{Z_{2}} \mapsto \overline{Z_{2}} \ , \ 
\overline{X_{3}} \mapsto \overline{X_{3}} \ , \nonumber \\
\overline{Z_{1}} \mapsto -\sin(\theta)\overline{Y_{1}} \, \overline{Z_{2}} \, \overline{X_{3}} +\cos(\theta)\overline{Z_{1}}\ , \nonumber \\
\overline{X_{2}} \mapsto \sin(\theta)\overline{X_{1}} \, \overline{Y_{2}} \,\overline{X_{3}} +\cos(\theta)\overline{X_{2}} \ , \nonumber\\ 
  \overline{Z_{3}} \mapsto -\sin(\theta)\overline{X_{1}} \, \overline{Z_{2}} \, \overline{Y_{3}} +\cos(\theta)\overline{Z_{3}}\ .
\end{align}

\begin{figure}[t]
    \centering
   \includegraphics[scale=0.5,keepaspectratio]{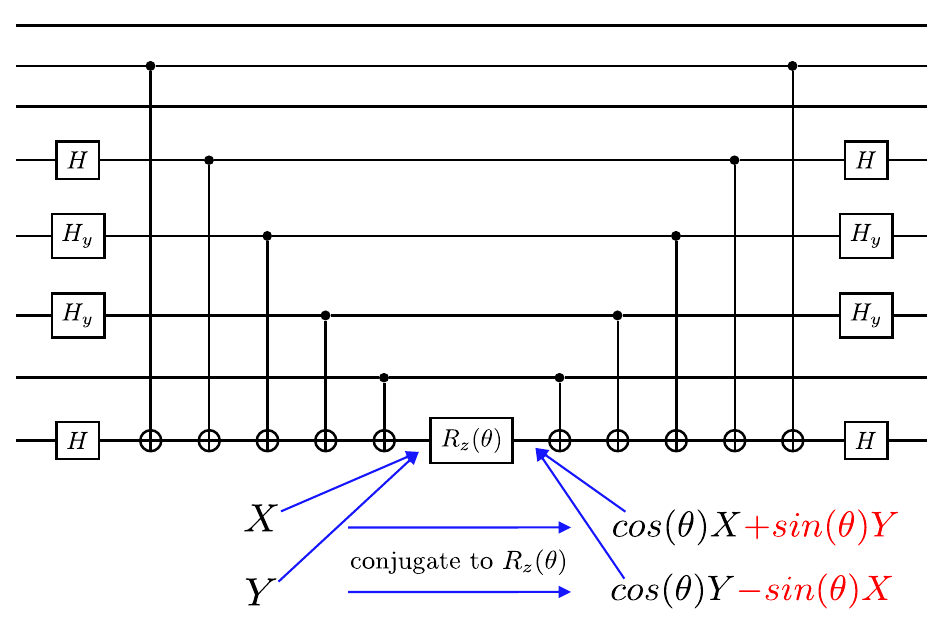}
    \caption{Physical realization of non-Clifford logical Trotter circuit in Fig.~\ref{fig:3_qubit_non_Clifford_logical} on $\llbr 8,3,3\rrbr$ code. The red contributions in the decomposition illustrate the Clifford-transformed part ($\sin(\theta)$-weighted term), while the black contributions ($\cos(\theta)$-weighted term) correspond to the unchanged input operator. This decomposition demonstrates the propagation of Pauli operators through the central $R_z(\theta)$ gate, maintaining the consistency of logical operator mappings.}
    \label{fig:8_qubit_non_Clifford_physical_logical_operators}
    \vspace*{-10pt}
\end{figure}

For the $\llbr 8,3,3 \rrbr$ code, the physical Trotter circuit is derived directly from the corresponding logical Clifford Trotter circuit, where the central Phase gate is replaced with $R_z(\theta)$ acting on the highest-index qubit within the support of the transvection. 
This new physical Trotter circuit is shown in Fig.~\ref{fig:8_qubit_non_Clifford_physical_logical_operators}.

In this specific case, logical operators are propagated according to this decomposition. For example, consider the logical operator 
\( Z_2 X_3 Z_5 X_8 \), which maps to 
\( X_3 X_4 X_5 Y_6 Z_7 \) in the Clifford case. 
The operator now maps as follows:
\begin{equation}
Z_2 X_3 Z_5 X_8 \mapsto \sin(\theta) X_3 X_4 X_5 Y_6 Z_7 + \cos(\theta) Z_2 X_3 Z_5 X_8,
\end{equation}
where the first term represents the contribution due to the Clifford gates and the second term corresponds to the unchanged input operator scaled by \( \cos(\theta) \). This decomposition is also shown in Fig.~\ref{fig:8_qubit_non_Clifford_physical_logical_operators}, illustrating how the symmetry of the Trotter circuit and the linearity of conjugation enable a straightforward propagation of Pauli operators through \( R_z(\theta) \). 

This example demonstrates how a non-Clifford logical Trotter circuit’s behavior remains consistent in the physical setting via the decomposition principle. 
Specifically, the replacement of the central gate with \( R_Z(\theta) \) preserves the operator propagation rules while introducing a precise weighting of the logical contributions by \( \sin(\theta) \) and \( \cos(\theta) \).


\begin{theorem}
\label{thm:non_Clifford_stabilizers}
For any non-Clifford logical Trotter circuit on any $\llbr n,k,d \rrbr$ stabilizer code, the stabilizer group is centralized by the physical Trotter circuit constructed in Theorem~\ref{thm:non_Clifford_logical_operators_propagation}.
\begin{IEEEproof}
\normalfont
According to Theorem~\ref{thm:Clifford any stabilizer code}, for the Clifford case, stabilizers are preserved and centralized by the logical Trotter circuit. This result directly applies to the Clifford physical circuit, which is derived from the logical Trotter circuit via symplectic transvection. Hence, the stabilizers remain centralized throughout the Clifford physical circuit.

The symmetry of the Trotter circuit ensures that stabilizers propagate consistently. As a stabilizer propagates from either the left or the right side of the Trotter circuit, it conjugates through each gate step-by-step. Due to the symmetry of the Trotter circuit, the resulting Pauli strings at corresponding steps on the front and back propagations are identical. This guarantees that the Pauli strings arriving at the central gate from the left and right sides are the same.

In the Clifford physical circuit, the central gate is a Phase gate. Since the Pauli strings reaching this gate are identical, the gate acting on the corresponding qubit must be \( Z \) or \( I \) to preserve the stabilizer structure.
Now, suppose we replace the central Phase gate with an arbitrary \( Z \)-rotation gate \( R_z(\theta) \). The symmetry of the Trotter circuit remains intact, and the propagation of Pauli strings continues to follow the same structure. The \( Z \)-rotation gate \( R_z(\theta) \) operates consistently with the \( Z \) or \( I \) requirement, as its effect on Pauli strings is an extension of the Phase gate's behavior. Therefore, stabilizers remain preserved. This is also applicable to trivial Pauli constraints in the non-Clifford case via a similar method.
\end{IEEEproof}
\end{theorem}

\begin{figure}[t]
    \centering
    \includegraphics[scale=0.45,keepaspectratio]{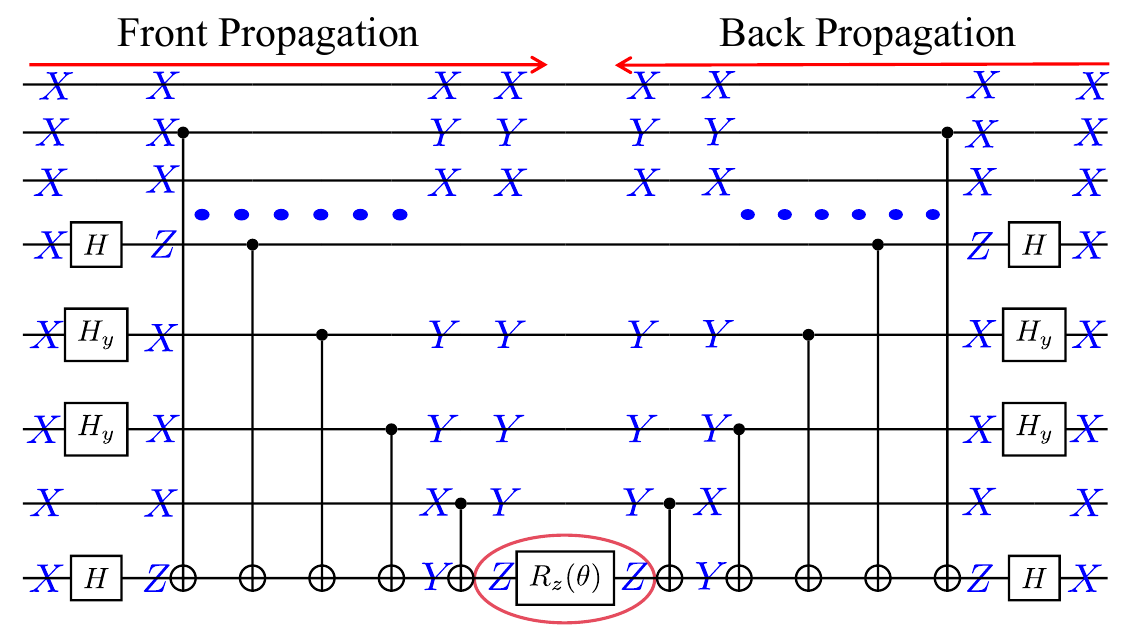}
    \caption{Propagation of the stabilizer $X_1X_2X_3X_4X_5X_6X_7X_8$
  through the Trotter circuit. Front and back propagation maintain symmetry, and the stabilizer remains identical throughout the circuit in the corresponding symmetric step. Propagation of Pauli operators in the intermediate parts are omitted.}
    \label{fig:8_qubit_stabilizer_propagation}
    \vspace*{-10pt}
\end{figure}

To demonstrate the preservation and centralization of stabilizers in the non-Clifford case, we again consider the $\llbr8,3,3\rrbr$ code, focusing on the stabilizer $S_{1}=X_1X_2X_3X_4X_5X_6X_7X_8$. 
The front and back propagation of the stabilizer is depicted in Fig.~\ref{fig:8_qubit_stabilizer_propagation}.
In the Clifford case, it has been established that stabilizers remain centralized throughout the circuit. This ensures that as the stabilizer $X_1X_2X_3X_4X_5X_6X_7X_8$ propagates through the front and back halves of the circuit, there is symmetry in the stabilizer operators at every step. 
This symmetry extends up to the central Phase gate, where the stabilizer operator on the left and right of the gate must remain consistent. 
At this central gate, only the qubit within the red circle in Fig.~\ref{fig:8_qubit_stabilizer_propagation} is affected. 
Due to the centralizing property of stabilizers, the operator on this qubit must be either $Z$ or $I$. 
In this example, the operator is $Z$, as determined by the structure of the stabilizer.
When the central Phase gate is replaced with $R_z(\theta)$, the stabilizer propagation remains unchanged.

Building upon the established results regarding stabilizers, we observe that the symmetry and centralizing properties of the Trotter circuit generalizes in a straightforward way. The preservation of stabilizers, even under the inclusion of non-Clifford gates such as $R_z(\theta)$, underscores the structured consistency of the circuit. This structured behavior not only ensures that stabilizers remain well-defined but also highlights a more general property of Pauli constraints in the circuit.
%
Specifically, we find that the product of input and output Pauli operators for non-trivial Pauli constraints remains constant, a property analogous to the Clifford case but extended to the non-Clifford regime. 
\begin{theorem}
\label{thm:non_Clifford_product}
 Let $h \coloneqq[a,b]\in\mathbb{F}_2^{2n}$.  
 For any non-Clifford Trotter circuit $U_h(\theta)$, 
the product of input and output operators for non-trivial Pauli constraints remains constant, given by 
\begin{equation}
\label{eq:non_Clifford_tranvection}
   U_h(-2\theta)=\exp(\imath \theta  E(a,b)). 
\end{equation}
\begin{IEEEproof}
\normalfont
The result can be obtained by a direct calculation of the conjugation and using trigonometric identities.
We omit the proof due to space constraints.
\end{IEEEproof}
\end{theorem}

Consider a Pauli error, $E(h')$, on the input qubits of an encoded Trotter circuit, $U_{\bar{h}}(\theta)$, which is, say, the residual of error correction done on a previous Trotter circuit using our framework.
If $\langle h', h \rangle_s = 0$, then the error remains unchanged at the end of the circuit.
However, if $\langle h', h \rangle_s = 1$, then the above result implies that the resulting error at the end of the circuit is $E(h')\, U_{\bar{h}}(-2\theta)$.
If we measure the syndrome, then it will be the same as that of $E(h')$, since $U_{\bar{h}}(-2\theta)$ commutes with stabilizers.
Therefore, if we get a nontrivial syndrome, then we can check if the error estimate, $\hat{E(h')}$, from the decoder anti-commutes with $E(\bar{h})$; if yes, then in addition to applying the correction, $\hat{E(h')}$, we must also apply a correction $U_{\bar{h}}(-2\theta)$.
If $\theta = \frac{\pi}{4}$, then this corresponds to a \emph{logical Clifford Trotter correction}.
Hence, the above structural insight provides a method to systematically track the effect of errors in the physical circuit.

\section{Fault Tolerant Trotter Circuits}
\label{sec:fault_tolerance}

To analyze fault-tolerance, express the rotation angle as $\theta/2 = \pi \sum_{i=1}^t x_i 2^{-t}$, where $x_i \in \mathbb{F}_2$, so that the finest angle involved is $\frac{\pi}{2^t}$.
We consider a concatenated coding strategy where each qubit of the $[[8,3,3]]$ outer code is further encoded into a ``$t$-orthogonal'' code~\cite{Bravyi-pra12,Haah-pra18} such as a color code with locality~\cite{Kubica-pra15} and single-shot property~\cite{Bombin-prx15,Kubica-natcomm22}.
However, we perform the inner encoding only after performing the single-qubit $H$ or $H_y$ gates physically on the outer code, followed by a round of error correction.
Then all the CNOT gates and the $R_Z$ rotation gate are performed transversally with error correction on the inner code after all gates.
Finally, we remove the inner code before performing the single-qubit $H$ or $H_y$ gates at the end physically on the $[[8,3,3]]$ code.
The next round of error correction would be performed after completing any single-qubit gates in the next Trotter stage.
If we replace the $[[8,3,3]]$ code with a good QLDPC code, then the inner code can be fixed with a small distance while the large distance of the QLDPC code exponentially suppresses logical errors.

Importantly, these error correction rounds are primarily removing entropy from the system. 
This is in contrast with the framework of logical Pauli measurements where error correction is critical to ensure the reliability of the measurements, which would otherwise have a domino effect on the remainder of the circuit.
The single-shot property or correlated decoding for transversal gates~\cite{Cain-prl24} can be leveraged to reduce the overhead of error correction.
Hence, this provides a promising path towards universal fault tolerance with Trotter circuits.

We also consider another approach to fault-tolerance by noting that a general quantum simulation algorithm iterates several Trotter circuits and ends with destructive $Z$-measurements.
We propose to use a concatenated QLDPC-cat DV-CV coding scheme~\cite{Ruiz-natcomm25} so that the system starts with a heavy bias towards $Z$-errors.
Then we can leverage bias-preserving CNOT gates in cat-based systems~\cite{Puri-sciadv20,Xu-prr22,Claes-prxq23} to preserve the $Z$-bias through the algorithm.
Note that the CNOT not only preserves the $Z$-bias in the qubits but also possesses a strong $Z$-bias in its own operation~\cite{Puri-sciadv20}.
We schedule syndrome measurements and error correction after each non-trivial Trotter circuit that retains the $Z$-bias.
It is clear that the non-Clifford Trotter blocks $R_{\overline{Z}}(\theta)$ will preserve the $Z$-bias at their outputs.
For the Clifford blocks, the single-qubit gates can alter the bias, so we must perform error correction before them to try and suppress $X$-errors.
At the end, the code qubits can all be individually measured in the $Z$-basis, since any logical $Z$ parity for the algorithm can be obtained from these destructive measurements.
Note that even for a non-CSS code such as the $[[8,3,3]]$ code, it is always possible to define logical $Z$ operators to be purely $Z$-type~\cite{Gottesman-phd97,Nielsen-book10}.
Therefore, any residual $Z$-errors, including logical ones, before the destructive measurements have no effect on the output distribution.
This establishes a path towards algorithmic fault-tolerance on a single code block in this framework.

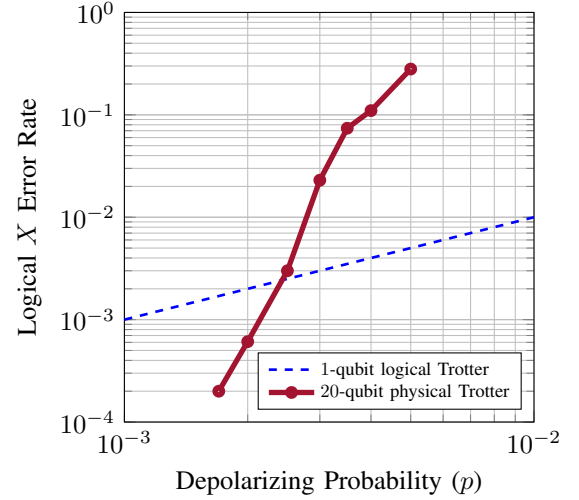
\begin{figure}
\centering

\begin{tikzpicture}[every node/.style={scale=1},scale=1]
\definecolor{mycolor1}{rgb}{0.63529,0.07843,0.18431}%

\begin{axis}[
    height=7cm,
    width=7cm,
    grid=both,
    xmode=log,
    ymode=log,
    xmin=1e-3,
    xmax=1e-2,
    ymin=1e-4,
    ymax=1,
    xlabel=  Depolarizing Probability ($p$), 
    ylabel=  Logical $X$ Error Rate,legend pos= south east,legend cell align=left,legend style={nodes={scale=0.7, transform shape}}
]

\addplot [color=blue,dashed,line width=1.0pt]
  table[row sep=crcr]{
  0.001  0.001\\
  0.01  0.01\\
  };\addlegendentry{$1$-qubit logical Trotter};

\addplot [color=mycolor1,solid,line width=2.0pt,mark size=1.4pt,mark=o,mark options={solid}]
  table[row sep=crcr]{
  0.005  0.28\\
  0.004  0.11\\
   0.0035 0.074\\
   0.003 0.023\\
  0.0025  3e-3\\
  0.002 6.1e-4\\
  0.0017 2e-4\\
  };\addlegendentry{$20$-qubit physical Trotter};
 
 \end{axis}
 \end{tikzpicture}
 
\caption{Simulation of $R_{\overline{Z}}(\frac{\pi}{2})$ on a $[[1054,140,20]]$ lifted product QLDPC code~\cite{Raveendran-quantum22} where the weight of $\overline{Z}$ is $d=20$. 
The plot shows the logical $X$ error rate at the output because these are the errors that would flip single-qubit $Z$-measurements if performed at the end of the algorithm. For comparison, the plot also shows the error rate on a single qubit if the $R_Z(\frac{\pi}{2})$ gate was performed without encoding the qubit. The pseudothreshold for this basic non-fault-tolerant setting with a non-tailored decoder is $\approx 2.5 \times 10^{-3}$, which is already higher than the error rates on, say, trapped-ions~\cite{H1-Datasheet}.}
\label{fig:LP_logical_S_bp-osd_dep_logicalX_error_rate}
\end{figure}


We conducted preliminary circuit-level noise simulations to test our framework and show the results in Fig.~\ref{fig:LP_logical_S_bp-osd_dep_logicalX_error_rate}.
We consider a $[[1054,140,20]]$ lifted product QLDPC code~\cite{Raveendran-quantum22} and implement $R_{\overline{Z}}(\frac{\pi}{2})$ on a \emph{targeted logical qubit} with logical $Z$ operator, $\overline{Z}$, of weight $d=20$.
This effectively performs a targeted logical phase gate on that logical qubit.
The physical circuit is a Clifford Trotter circuit on $d=20$ qubits with no $H$ or $H_y$ gates.
We set idling qubits and CNOTs to have a single-qubit and two-qubit depolarizing channel, respectively, with the same error rate $p$.
After the Trotter circuit, an ideal round of syndrome extraction is performed and the error is estimated using a plain BP-OSD decoder~\cite{Panteleev-quantum21}.
We observe a pseudo-threshold of $\approx 2.5 \times 10^{-3}$, which is quite promising because the circuit is not fault-tolerant, the CNOTs fail with the same rate as idling qubits, there is no bias in the noise, and the decoder is not tailored to the error propagation.
Moreover, this pseudo-threshold is already achievable in current hardware, e.g., trapped-ions~\cite{H1-Datasheet}.
Hence, these initial results show that our framework can indeed lead to simple, low-overhead, fault tolerant quantum simulation on a single code block.

\section{Conclusion}
\label{sec:conclusion}

In this work, we established an exciting new path towards fault-tolerant quantum simulation by synthesizing logical Trotter circuits \emph{as a whole}.
This disrupts the conventional wisdom of synthesizing logical circuits gate by gate.
In future work, we will systematically analyze the performance of this approach for quantum simulation and rigorously compare with existing methods that use lattice surgery and magic state distillation.

\appendices








\end{document}